%% file: bare_conf.tex
\documentclass[conference]{IEEEtran}
\usepackage{ifpdf}

\usepackage{cite}

\ifCLASSINFOpdf
\usepackage[pdftex]{graphicx}
\else
   \usepackage[dvips]{graphicx}
\fi
\usepackage{amsmath}
\usepackage{algorithmic}

\usepackage{array}

\ifCLASSOPTIONcompsoc
  \usepackage[caption=false,font=normalsize,labelfont=sf,textfont=sf]{subfig}
\else
  \usepackage[caption=false,font=footnotesize]{subfig}
\fi
\usepackage{fixltx2e}

\usepackage{stfloats}
\usepackage{url}

\input{preamble/packages.tex}

\input{preamble/commands.tex}

\input{preamble/acronyms.tex}

\makeglossaries

\begin{document}
%
\title{ColibriES: A Milliwatts RISC-V Based Embedded System Leveraging Neuromorphic and Neural Networks Hardware Accelerators for Low-Latency Closed-loop Control Applications}

\author{\IEEEauthorblockN{Georg Rutishauser\IEEEauthorrefmark{1}, Robin Hunziker\IEEEauthorrefmark{1}, Alfio Di Mauro\IEEEauthorrefmark{1}, Sizhen Bian\IEEEauthorrefmark{1}, Luca Benini\IEEEauthorrefmark{1}\IEEEauthorrefmark{2}, Michele Magno\IEEEauthorrefmark{1}}
\IEEEauthorblockA{\IEEEauthorrefmark{1}\textit{Departement Informationstechnologie und Elektrotechnik, ETH
    Z{\"u}rich, Switzerland}}
\IEEEauthorblockA{\IEEEauthorrefmark{2}\textit{Dipartimento di Ingegneria dell'Energia Elettrica e dell'Informazione, Universit\`{a} di Bologna, Bologna, Italy}}
}
\maketitle

\begin{abstract}
End-to-end event-based computation has the potential to push the envelope in latency and energy efficiency for edge AI applications. Unfortunately, event-based sensors (e.g., DVS cameras) and neuromorphic spike-based processors (e.g., Loihi) have been designed in a decoupled fashion, thereby missing major streamlining opportunities. 
This paper presents ColibriES, the first-ever neuromorphic hardware embedded system platform with dedicated event-sensor interfaces and full processing pipelines. ColibriES includes event and frame interfaces and data processing, aiming at efficient and long-life embedded systems in edge scenarios. ColibriES is based on the Kraken system-on-chip and contains a heterogeneous parallel ultra-low power (PULP) processor, frame-based and event-based camera interfaces, and two hardware accelerators for the computation of both event-based spiking neural networks and frame-based ternary convolutional neural networks. This paper explores and accurately evaluates the performance of event data processing on the example of gesture recognition on ColibriES, as the first step of full-system evaluation. In our experiments, we demonstrate a chip energy consumption of 7.7 \si{\milli\joule} and latency of 164.5 \si{\milli\second} of each inference with the DVS Gesture event data set as an example for closed-loop data processing, showcasing the potential of ColibriES for battery-powered applications such as wearable devices and UAVs that require low-latency closed-loop control.
\end{abstract}


%
\IEEEpeerreviewmaketitle

\section{Introduction}
Neuromorphic computing is a promising paradigm for low latency, energy-efficient artificial intelligence \cite{nwakanma2021edge} for a wide range of embedded applications \cite{vitale2022neuromorphic}. Complementing this novel processing paradigm, event-based sensors have the potential to unlock the full potential of neuromorphic hardware for applications such as intelligent sensing and ultra-fast robotics \cite{lele2021end}. On the other hand,  frame-based processing using \glspl{DNN} is currently offering impressive performance and is being heavily tuned for low-power, and energy-efficient embedded platforms \cite{huynh2022implementing}. 

Combining the advantages of event-based image sensors, neuromorphic processing, traditional image sensors, and deep neural networks yields hybrid hardware and software systems that can outperform the current state of the art in low-latency, energy-efficient systems like unmanned vehicles. For instance, optical flow estimation is one of the most used methods for autonomous navigation. Improving its speed and accuracy will improve the entire autonomous vehicles field \cite{akbari2021applications}. Our work aims at building the first hybrid system combining frame information with \gls{DNN}-based processing and event information with SNN-based processing for autonomous vehicles and robots \cite{varshika2022design}.

\subsection{Motivation}
\begin{table*}[htbp]
      \caption{Comparison of ColibriES to recent work}
      \label{tbl:rel_works_comparison}
    \begin{threeparttable}
      \begin{tabularx}{\textwidth}{X|r|r|r|r|r|r}
        & \cite{ref:benchmark_kws_neuromorphic} & \cite{ref:kws_truenorth} &
        \cite{ref:dvs_emg_gesture_benchmark} & \cite{ref:dvs_loihi} &
        \cite{ref:dvs_truenorth} & \textbf{ColibriES (ours)} \\\hline
        Platform & Loihi & TrueNorth & Loihi & Loihi & TrueNorth & Kraken (SNE)
        \\
        Application \tnote{a} \ & \acrshort{kws} & \acrshort{kws} & GR                         & GR
                                & GR             & GR \\
        Input Data              & MFCC \tnote{d} & Audio          & DVS + sEMG Evts. \tnote{d} & DVS Evts. & DVS Evts.
                                & DVS Evts. \\
        Dataset                 & Custom         & TIDIGITS       & Custom                     & DVS128    & DVS128   & DVS128 \\
        \# of Classes           & 2              & 11             & 5                          & 11        & 11       & 11 \\
        Network \tnote{b}                 & 2L SMLP        & 7L SCNN        & 6L SCNN + 4L SMLP          & 5L SCNN   & 16L SCNN & 
        7L SCNN \\
        Accuracy                                      & $95.9\%$               & $90.8\%$-$95.1\%$ \tnote{e}         & $96\%$      & $90.5\%$ & $86.5\%$-$94.6\%$ \tnote{e} & $83\%$ \\
        End-to-End?                                   & \xmark                 & \cmark                              & \xmark      & \xmark   & \cmark                      & \cmark \\
        $P_{proc, inf}$ (\si{\milli\watt}) \tnote{b}  & 110                    & 14.4
        (85)-38.6 (91) \tnote{e}                               &
        141.9                                         & N/A                    & 88.5 - 178.8                        & 35.6 \\
        $P_{proc, idle}$ (\si{\milli\watt}) \tnote{b} & 29.2                     & 12.1 (82.7) - 30.3 (82.7) \tnote{e} & 29.2
        \cite{ref:benchmark_kws_neuromorphic}         &  29.2
        \cite{ref:benchmark_kws_neuromorphic}         & 68.8 - 134.4 \tnote{f} & 17.7 \\
        $E_{proc, inf}$ (\si{\milli\joule}) \tnote{d} & 0.371                  &
        15.9 (93.6) - 42.5 (100.1) \tnote{g}
                                                      & 5.9            & N/A                                 & 29.8 & \textbf{7.7} \\
      \end{tabularx}
      \begin{tablenotes}
        \item[a] \textbf{KWS:} Keyword Spotting, \textbf{GR:} Gesture
          Recognition
          \item[b] $\bm{N}$\textbf{L SCNN:} $N$-layer spiking convolutional neural network, $\bm{N}$\textbf{L SMLP:} $N$-layer spiking multi-layer perceptron
          \item[c] $P_{proc, \{inf, idle\}}$: Processor chip/full system
            chip power during inference/while idle
          \item[d] $E_{proc, inf}$: Processor chip/full system
            energy per inference. Energy normalized to 6 inf/s for GR.
          \item[e] Preprocessing not included in energy measurements.
          \item[f] Lowest-power and highest-accuracy implementations.
            Numbers in parentheses are energy/power omit scaling for resource
            usage.
          \item[g] We assume the preprocessing and classification pipeline needs
            to run for the entire length of a sample (\SI{1104}{\milli\second}).
      \end{tablenotes}
      \end{threeparttable}

\end{table*}

Most current systems using visual sensing to perform intelligent tasks rely on frame-based cameras \cite{chai2021deep}. Conventional video sensors record the entire image at a fixed rate and resolution. These vision systems use sequences of RGB frames, which are heavily redundant and inefficient to process. Event-based cameras (silicon retinas) have introduced a new vision paradigm by mimicking the biological retina. Instead of measuring the intensity of every pixel at fixed time intervals, they report events of significant pixel intensity changes. Each such event is represented by its position, the sign of change, and timestamp in microsecond resolution. Because of their asynchronous operation principle, they are a natural match for \glspl{SNN} \cite{steffen2019neuromorphic} rather than \glspl{DNN}. 

State-of-the-art \gls{DNN} approaches in machine learning provide excellent results for vision tasks with standard cameras.
In contrast, asynchronous event sequences require special handling, and \glspl{SNN} can take advantage of this asynchronicity, offering very low detection latency in the order of microseconds \cite{pfeiffer2022visual}. However, \gls{SNN}-based algorithms are still evolving, and current \glspl{SNN} are still not comparable in terms of accuracy to \gls{DNN}-based approaches. 

Another aspect limiting the employment of energy-proportional event-based cameras in real application scenarios requiring ultra-low-power devices is the high power consumption spent at such sensors' interfaces. Event cameras often expose non-standard communication protocols, and data acquisition is commonly carried out with a high-power \gls{FPGA}, and a USB interface \cite{schiavone2021arnold}. 

Neuromorphic processors execute \glspl{SNN} on large-scale integrated circuits \cite{qiu2021novel} by mimicking the natural biological structures of the nervous
system, implementing neurons and synapses directly in silicon.
A typical neuromorphic processor can simulate hundreds of thousands or even millions of spiking neurons in real time. However, to achieve truly power-efficient real-time operation, such platforms should also be equipped with microcontroller subsystems capable of performing actuation and closed-loop control. Even advanced prototypes, e.g., the Loihi Kapoho Bay board, do not present such features as they are generally conceived as research and evaluation testbeds. Kraken, a novel RISC-V-based \SOC addressing this challenge, has been recently proposed \cite{di2022sne}.   Kraken includes a heterogeneous parallel ultra-low power (PULP) processor \cite{wang2020fann} and two hardware accelerators for both spiking neural networks and frame-based convolutional neural networks. A key feature of the Kraken \SOC is that both RGB camera and event-based camera interfaces are present, as well as the general-purpose input output pins useful for direct control of motors and other devices. Kraken can deliver up to 2TOP/s/W 2bit SIMD on the RISC-V cluster and 1.1TSynOP/s/W on the neuromorphic accelerator, which makes it suitable for embedded battery-operated devices requiring low-latency closed-loop control. \Cref{tbl:rel_works_comparison} compares our proposed ColibriES system with previous works implementing advanced tasks on neuromorphic processors. The only published works implementing true end-to-end neuromorphic applications are based on IBM's TrueNorth platform \cite{ref:truenorth_10y},  a purely neuromorphic processor that lacks the advanced compute and control capabilities enabled by the Kraken \SOC, relying solely on \glspl{SNN} for all processing tasks, which in turn restricts the range of tasks that can be solved on the platform.

\subsection{Contribution}
\begin{figure}[!b]
    \centerline{\includegraphics[width=\linewidth]{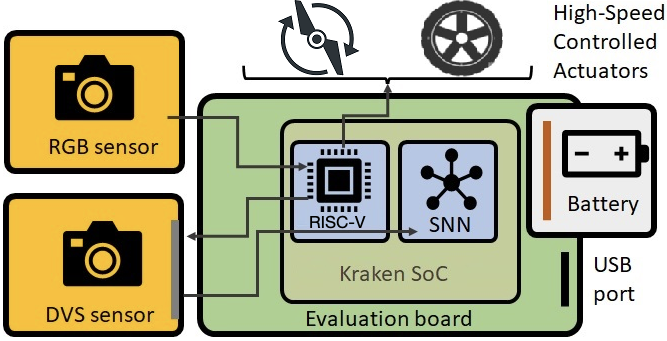}}
    \caption{Architecture of the embedded platform developed as evaluation board hosting the novel Kraken SoC, which includes PWM pins for high-speed control.}
    \label{fig:HighLevel}
\end{figure}

This paper presents ColibriES, the first-ever neuromorphic truly embedded evaluation platform hosting the Kraken \SOC. The ColibriES board was designed to accurately measure the complete processing pipeline's power consumption, energy, and latency. ColibriES can connect to both event- and frame-based cameras that directly interface with the Kraken \gls{SOC}. The platform is designed for battery-powered applications requiring low-latency closed-loop control feedback, such as wearable electronics, \glspl{UAV}, and other applications that impose stringent requirements on the computational and energy resources available to perform the processing. \Cref{fig:HighLevel} shows the high-level architecture of the proposed platform and provides an overview of how the system could be used as a high-speed closed-loop controller.
Evaluations of power consumption, functionality, and energy efficiency are presented with experimental measurements, demonstrating the versatility and low power of ColibriES. 

The rest of this paper is organized as follows: Section \ref{section_2} describes the proposed ColibriES platform with a particular focus on Kraken's architectural features. Section \ref{section_3} shows the experimental results, and section \ref{section_4} concludes the paper.

\section{System Architecture}
\label{section_2}

In this section, we present the heart of ColibriES, the heterogeneous Kraken \SOC, and the evaluation board used in our experiments.

\subsection{Kraken \SOC}

\begin{figure}[!t]
    \centerline{\includegraphics[width=1\linewidth]{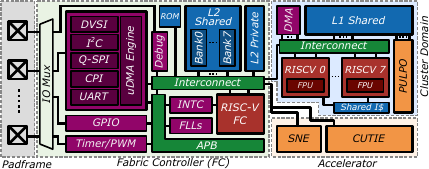}}
    \caption{Kraken \SOC block diagram}
    \label{fig:krakenarchi}
\end{figure}
The Kraken \SOC has a heterogeneous architecture composed of three main subsystems. Fig. \ref{fig:krakenarchi} shows a block diagram representation of the Kraken chip. The first subsystem, the \FC, is built around a 32-bit RISC-V core which acts as the central programmable control unit for the whole \SOC. The \FC hosts the main interconnection busses to the main L2 memory and the \APB, which controls all the \SOC peripherals. The \FC domain also contains a full set of on-chip peripherals such as a RISC-V compliant debug unit, four programmable timer-counters and a power management unit.

The second subsystem is a RISC-V-based general-purpose accelerator called the \textit{cluster}. The cluster domain hosts eight RISC-V cores enhanced with specialized \ISA extensions like hardware loops, multiply and accumulate, and vectorial instructions for low-precision \ML workloads. Each RISC-V core of the cluster accelerator is equipped with a private \FPU and implements the full 32-bit RISC-V floating-point \ISA extension.
The cluster also features a \SI{128} Kb L1 \TCDM. The L1 memory can serve memory requests from all cluster cores in parallel with single-cycle access latency and a low average contention rate (\textless 10\% even on the most data-intensive kernels). Fast event management, parallel thread dispatching, and synchronization are supported by a dedicated hardware block (HW Sync), enabling very fine-grained parallelism and high energy efficiency in parallel workloads. The cluster can be clock-gated at single-core granularity, reducing the dynamic power consumption while waiting for core synchronization.

The third domain, the \EHWPE, hosts two accelerators. One is the \SNE neuromorphic accelerator presented in \cite{di2022sne}, and the other is CUTIE, a ternary weight neural network accelerator presented in \cite{CUTIE}. The \EHWPE domain operates on two independent clock domains. One clock is generated by the \FC \FLL, which is used to clock the interface logic at the boundary between the accelerators and the main system interconnect. The other clock is generated by a dedicated \EHWPE \FLL, which is used to clock the accelerator's computing engines. Like many peripherals of the \SOC, the accelerators are programmed via memory-mapped register interfaces. 
The \FC domain hosts a vast set of IO peripherals commonly adopted on microcontroller-like \SOCs. The chip features 76 digital pads, 44 of which can be used to arbitrarily map any pin of the peripherals mentioned above to a chip pad in an all-to-all crossbar configuration, providing high IO mapping flexibility. The pin mapping is configurable via a memory-mapped register interface connected to the \APB. Each peripheral can generate interrupts depending on data transmission events; such events can then be routed to the interrupt controller of the RISC-V core.

An autonomous IO subsystem, the \textit{{\textmu}DMA}\cite{pullini2019mr},  hosts all the IO peripherals. This system can be programmed to autonomously orchestrate data transfers from the peripherals to the L2 memory and vice versa, freeing the RISC-V core from micro-managing such data transfers.
Moreover, Kraken's  {\textmu}DMA is equipped with a dedicated hardware interface for the ultra-low power \DVS presented in \cite{li2019132}. Such an interface was presented in \cite{flyDVSDiMauro} and allows streaming events directly from the event camera to Kraken.

To carefully control the chip's overall power consumption and enable a versatile application-specific power management scheme, Kraken is subdivided into three independent power domains. Such power domain partitioning allows to switch-off of unused parts of the system that the running application might not require. The power-up/down of the controllable domains can be performed at run time. Figure \ref{fig:krakenarchi} also shows the chip power domain partitioning, \SOC, Cluster, and \EHWPE, respectively. The \FC domain is always on because it hosts essential components needed for \SOC management tasks such as the boot procedure, clock generator initialization, or accelerator reset and programming.

\subsection{Kraken evaluation board}
\begin{figure}[!t]
    \centerline{\includegraphics[width=1\linewidth]{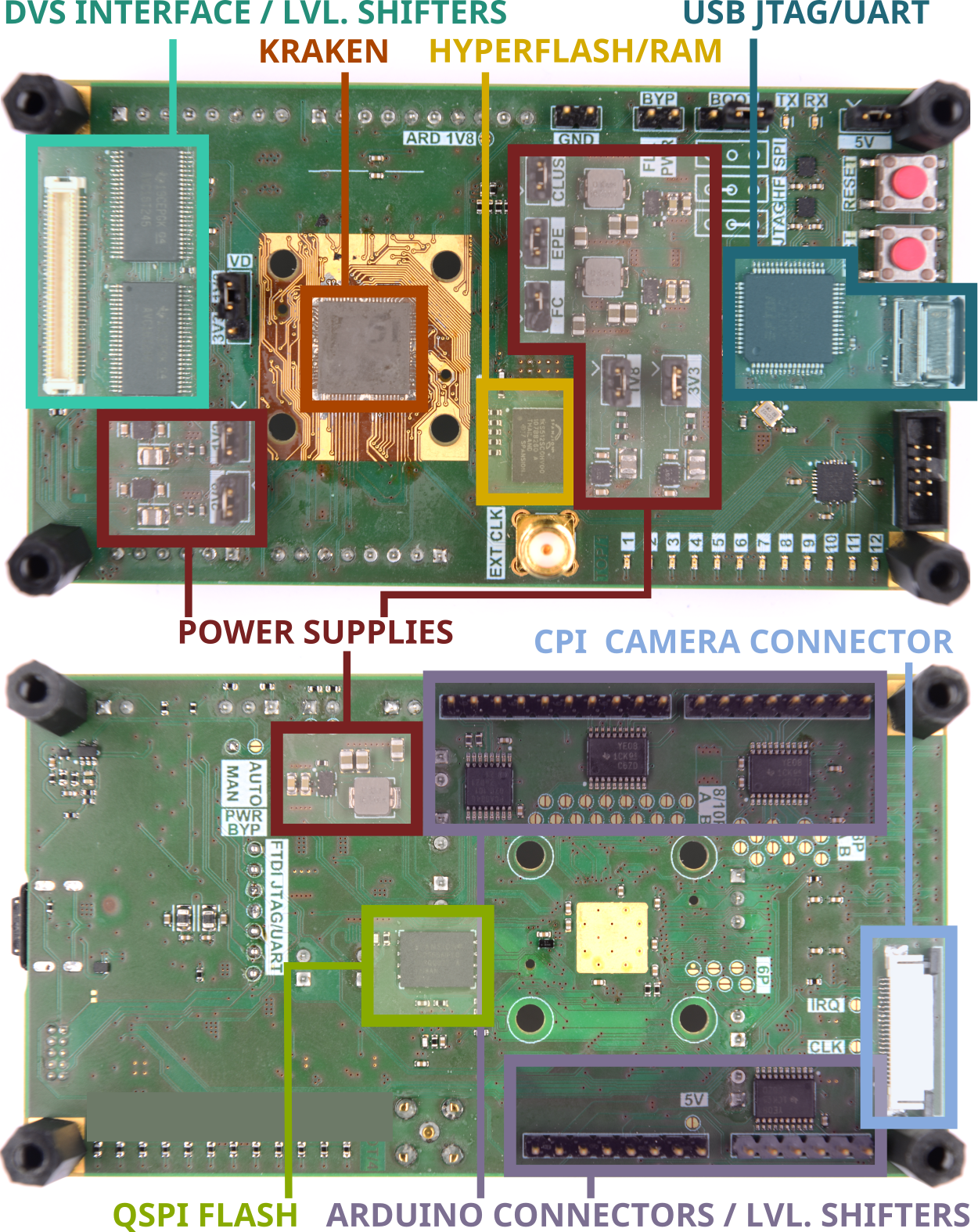}}
    \caption{Kraken evaluation board}
    \label{fig:krakenpcb}
\end{figure}

The Kraken evaluation board integrates the Kraken chip with the external
components needed to implement complete applications. A USB-C connector is used
for JTAG and UART data transfer and supplies
a \SI{5}{\volt} rail from which all other power supplies are derived. Kraken's
three power domains are supplied by individually runtime-configurable buck
converters, allowing for application-controlled \gls{DVFS}. Connectivity to the
\gls{DVS} camera is provided through an 80-pin low-profile board-to-board
connector, which also supplies the camera board with the required \SI{1.2}{\volt}
and \SI{5}{\volt} rails. Off-chip memory is present in the form of a combined
HyperFlash/HyperRAM chip and a quad-SPI flash memory chip, which can also be used to store application code for standalone booting. Arduino headers and a
\gls{CPI} ribbon connector provide additional connectivity, with level shifters
between every off-chip connector and Kraken's I/O pins. An annotated photograph
of the board is shown in Fig. \ref{fig:krakenpcb}.

\section{Evaluation}
\label{section_3}

\begin{table}[htb]
      \caption{DVS Gesture Network Parameters}
      \label{tbl:gesture_network_loihi}
    \begin{threeparttable}
      \begin{tabularx}{\textwidth}{r|r|r|r|r|r}
        & Type & Size & Feature Size & Features & Stride 
        \\
        \cline{1-6}
        0 & Input & 128x128x2 & - & - & - 
        \\
        1 & Pool & 32x32x2 & 4x4x1 & 2 & 4 
        \\
        2 & Conv & 32x32x16 & 3x3x2 & 16 & 1 
        \\
        3 & Pool & 16x16x16 & 2x2x1 & 16 & 2 
        \\
        4 & Conv & 16x16x32 & 3x3x1 & 32 & 1 
        \\
        5 & Pool & 8x8x32 & 2x2x1 & 32 & 2 
        \\
        6 & Full & 512 & 2048 & 512 & - 
        \\
        7 & Full & 11 & 512 & 11 & - 
        \end{tabularx}
      \end{threeparttable}

\end{table}

In this section, we evaluate the end-to-end performance of ColibriES when executing SNN inference as the first step of the full-system evaluation. We used the IBM DVS Gesture event data set \cite{amir2017low}, which contains 11 hand gestures from 29 subjects under three illumination conditions from a DVS128 event camera. Six subjects' data are used for testing and the remaining for training. The \gls{SNN} used for our experiments is reported in Table \ref{tbl:gesture_network_loihi}. The network is composed of two convolutional layers and two fully-connected layers. To train the network, we used a supervised learning approach based on back-propagation, derived from the one presented in \cite{wu2018spatio}, which has been implemented in PyTorch. In this framework, the Neuron dynamic of the \LIF neurons implemented in SNE is accurately modeled to closely reflect the hardware implementation and minimize the mismatch between the accuracy achieved during training and the one achieved on the real hardware platform.

We break down the closed loop into three distinct parts: data acquisition on the \FC through the dedicated DVS interface, data processing on the engines, which includes a spike preprocessing step in the cluster and a spike train inference step in the SNE, and actuators control using PWM signals. For advanced processing tasks, neural networks that exceed \gls{SNE}'s output neuron capacity are executed on the accelerator in a tiled way, and the SNE is used in a time-domain-multiplexing fashion. The preprocessing step performed on the cluster is necessary to assemble a single input event stream from multiple output tiles and create the tiled input streams for the tiles of the successive layer.
Note that the main system clock of Kraken can vary from a few \si{\kilo\hertz} to \SI{200}{\mega\hertz}, and the typical operating frequency is \SI{50}{\mega\hertz}. In typical operating conditions, the latency of changing an output actuation signal like a PWM signal is a few system clock cycles, i.e., below \SI{1}{\micro \second}. Therefore, we considered this latency negligible compared to data acquisition and processing.  
\Cref{tbl:benchmarkingColibriES} reports the energy and latency metrics of the ColibriES system for the gesture recognition task execution. Using a 300 milliseconds window input, Kraken requires 164.5 \si{\milli\second} and 7.7 \si{\milli\joule} to perform a DVS-to-label prediction, meeting real-time processing constraints. The inference execution on SNE takes only 32 \si{\milli\second} and 1.4 \si{\milli\joule}. 

\begin{table}[!t]
\caption{Performance metrics of ColibriES on DVS-HandGesture Dataset with accuracy of 83\%}
\label{tbl:benchmarkingColibriES}
\centering
\begin{threeparttable}
\begin{tabularx}{\linewidth}{Xrrrr}
\hline
\multirow{2}{*}{Proc. Stage} & \multirow{2}{*}{Time (ms)} & \multicolumn{2}{c}{Power (\si{\milli\watt})\tnote{a}} & \multirow{2}{*}{Energy (\si{\milli\joule})\tnote{c}}\\\cline{3-4}
& & Idle & Active & \\\hline
Data Acquisition (\FC) &  1.5 &  3.5 & 3.8 & 0.006  \\ \hline
Preprocessing (Cluster)\tnote{c}, & 131 &  6.5 & 34 & 4.6 \\\hline
\SNN Inference (\SNE) & 32 & 7.7  & 44 & 1.4 \\
\hline
\textbf{Total} & 164.5 & 17.7 & 35.6 \tnote{c} & \textbf{7.7}\\\hline
\end{tabularx}
\begin{tablenotes}
    \item All power numbers measured at $V_{DD}=\SI{0.65}{\volt}$
    \item[a] Idle power measured with respective components clock-gated.
    \item[b] Total energy is computed as the sum of active energy contributions and idle energy of inactive components.
    \item[c] Average total power consumption during inference
\end{tablenotes}
\end{threeparttable}
\end{table}

%
%
%
%
%
%

\section{Conclusion}
\label{section_4}
This paper presented ColibriES, an energy-efficient heterogeneous platform including an event-based camera interface and an RGB interface, together with the control features of a standard microcontroller. The platform core is the Kraken system-on-chip, a PULP processor with eight RISC-V cores and two dedicated hardware accelerators for machine learning: a neuromorphic processor and a frame base machine learning accelerator. As the first step of system evaluation on ColibriES, this paper reports the evaluation of end-to-end gesture recognition from event camera data with the dedicated DVS interface and the neuromorphic core. Thanks to the direct access to the stream of events from the event camera, ColibriES can achieve real-time event processing by executing SNN inference on sensor data. Experimental results show that our platform can perform low-latency tasks under stringent power constraints, and it is suitable for battery-powered operation and low-latency applications. On the IBM DVS Gesture event data set, ColibriES achieves real-time 164ms processing on 300ms time window samples, consuming only 7.7mJ of total energy and meeting the severe latency and energy constraints of emerging high-speed closed-loop control applications, such as autonomous drones and wearable control embedded systems, while keeping the power in the mW range.


\section*{Acknowledgment}
The authors would like to thank \textit{armasuisse Science \& Technology} for funding this research.

\newpage



%

\bibliography{library_georgr}




\end{document}

%% file: preamble/packages.tex
\usepackage{cite} 

\usepackage{amsmath,amssymb,amsfonts}
\usepackage{algorithmic}
\usepackage{graphicx}
\usepackage{textcomp}
\usepackage{xcolor}
\usepackage{threeparttable}
\usepackage{tabularx}
\usepackage{url}
\usepackage{siunitx}
\usepackage{caption}
\usepackage{cleveref}
\usepackage{xspace} 
\usepackage[acronym]{glossaries}  
\usepackage{pifont}
\usepackage{ulem}
\usepackage{textgreek}
\usepackage{multirow}
\usepackage{comment}
\usepackage{bm}

%% file: preamble/commands.tex
\newcommand{\cmark}{\ding{51}}%
\newcommand{\xmark}{\ding{55}}%
\glsdisablehyper

%% file: preamble/acronyms.tex
\newcommand{\myacronymentry}[4]{
\newglossaryentry{#1}{
  type = \glsdefaulttype,
  name = #2,
  short = #2,
  description = #3,
  first = #3 (#2),
  text = #2,
  plural = #2\glspluralsuffix,
  firstplural = #3\glspluralsuffix\space (#2\glspluralsuffix),
  sort = #4}
}

\myacronymentry{ADC}{ADC}{analog-to-digital converter}{adc}

\myacronymentry{AGC}{AGC}{automatic gain control}{agc}

\myacronymentry{ASIC}{ASIC}{application-specific integrated circuit}{asic}

\myacronymentry{BER}{BER}{bit error rate}{ber}

\myacronymentry{BPSK}{BPSK}{binary phase-shift keying}{bpsk}

\myacronymentry{CMOS}{CMOS}{complementary metal-oxide semiconductor}{cmos}

\myacronymentry{CP}{CP}{cyclic prefix}{cp}

\myacronymentry{DAB}{DAB}{digital audio broadcasting}{dab}

\myacronymentry{DAC}{DAC}{digital-to-analog converter}{dac}

\myacronymentry{DC}{DC}{direct current}{dc}

\myacronymentry{DFT}{DFT}{discrete Fourier transform}{dft}

\myacronymentry{DSL}{DSL}{digital subscriber line}{dsl}

\myacronymentry{FDD}{FDD}{frequency-division duplex}{fdd}

\myacronymentry{FFT}{FFT}{fast Fourier transform}{fft}

\myacronymentry{FPGA}{FPGA}{field-programmable gate array}{fpga}

\myacronymentry{HDTV}{HDTV}{high-definition television}{hdtv}

\myacronymentry{IDFT}{IDFT}{inverse discrete Fourier transform}{idft}

\myacronymentry{IEEE}{IEEE}{institute of electrical and electronics engineers}{ieee}

\myacronymentry{IFFT}{IFFT}{inverse fast Fourier transform}{ifft}

\myacronymentry{ISI}{ISI}{inter-symbol interference}{isi}

\myacronymentry{LAN}{LAN}{local area network}{lan}

\myacronymentry{LTE}{LTE}{long term evolution}{lte}

\myacronymentry{MIMO}{MIMO}{multiple-input multiple-output}{mimo}

\myacronymentry{MMSE}{MMSE}{minimum mean squared error}{mmse}

\myacronymentry{OFDM}{OFDM}{orthogonal frequency-division multiplexing}{ofdm}

\myacronymentry{PAPR}{PAPR}{peak-to-average power ratio}{papr}

\myacronymentry{PHY}{PHY}{physical layer}{phy}

\myacronymentry{QAM}{QAM}{quadrature amplitude modulation}{qam}

\myacronymentry{QPSK}{QPSK}{quadrature phase-shift keying}{qpsk}

\myacronymentry{QoS}{QoS}{quality of service}{qos}

\myacronymentry{RF}{RF}{radio frequency}{rf}

\myacronymentry{RSSI}{RSSI}{received signal strength indicator}{rssi}

\myacronymentry{SISO}{SISO}{single-input single-output}{siso}

\myacronymentry{SNR}{SNR}{signal-to-noise ratio}{snr}

\myacronymentry{SQRD}{SQRD}{sorted QR decomposition}{sqrd}

\myacronymentry{ST}{ST}{space-time}{st}

\myacronymentry{STBC}{STBC}{space-time block code}{stbc}

\myacronymentry{TDD}{TDD}{time-division duplex}{tdd}

\myacronymentry{UMTS}{UMTS}{universal mobile telecommunications system}{umts}

\myacronymentry{UMC}{UMC}{United Microelectronics Corporation}{umc}

\myacronymentry{VLSI}{VLSI}{very large scale integration}{vlsi}

\myacronymentry{WiMAX}{WiMAX}{worldwide interoperability for microwave access}{wimax}

\myacronymentry{WLAN}{WLAN}{wireless local area network}{wlan}

\myacronymentry{WMAN}{WMAN}{wireless metropolitan area network}{wman}

\myacronymentry{DVS}{DVS}{dynamic visual sensor}{dvs}
\newcommand{\DVS}{\gls{DVS}\xspace}

\myacronymentry{DVSI}{DVSI}{dynamic visual sensor interface}{dvsi}

\myacronymentry{DASI}{DASI}{dynamic audio sensor interface}{dasi}

\myacronymentry{LUT}{LUT}{look up table}{lut}

\myacronymentry{QSPI}{QSPI}{quad-serial peripheral interface}{qspi}

\myacronymentry{IIC}{I$^2$C}{Inter-Integrated Circuit}{i2c}

\myacronymentry{IIS}{I$^2$S}{Inter-Integrated-Circuit Sound}{i2s}

\myacronymentry{MCU}{MCU}{micro-controller unit}{mcu}

\myacronymentry{SOC}{SoC}{system-on-chip}{soc}
\newcommand{\SOC}{\gls{SOC}\xspace}
\newcommand{\SOCs}{\glspl{SOC}\xspace}

\myacronymentry{SOA}{SoA}{state-of-the-art}{soa}

\myacronymentry{STA}{STA}{static timing analysis}{sta}

\myacronymentry{AER}{AER}{Address Event Representation}{aer}

\myacronymentry{AETR}{AETR}{Address-Event-Timestamp Representation}{aetr}

\myacronymentry{NRZ}{NRZ}{Non Return to Zero}{nrz}

\myacronymentry{PPA}{PPA}{power performance analysis}{ppa}

\myacronymentry{SNE}{SNE}{sparse neural engine}{sne}
\newcommand{\SNE}{\gls{SNE}\xspace}

\myacronymentry{DMA}{DMA}{direct memory access}{dma}

\myacronymentry{ALIF}{ALIF}{adaptive leaky-integrate and fire}{alif}

\myacronymentry{LIF}{LIF}{leaky-integrate and fire}{lif}
\newcommand{\LIF}{\gls{LIF}\xspace}

\myacronymentry{FIFO}{FIFO}{first-in first-out}{fifo}

\myacronymentry{SNN}{SNN}{spiking neural network}{snn}
\newcommand{\SNN}{\gls{SNN}\xspace}

\myacronymentry{DNN}{DNN}{deep neural network}{dnn}

\myacronymentry{ANN}{ANN}{artificial neural network}{ann}

\myacronymentry{BNN}{BNN}{binary neural network}{bnn}

\myacronymentry{CDC}{CDC}{clock domain crossing}{cdc}

\myacronymentry{IOT}{IOT}{internet of things}{iot}

\myacronymentry{NTC}{NTC}{near-threshold computing}{ntc}

\myacronymentry{SOP}{SOP}{synaptic operation}{sop}

\myacronymentry{GPU}{GPU}{graphics process unit}{gpu}

\myacronymentry{CPU}{CPU}{central process unit}{cpu}

\myacronymentry{TPU}{TPU}{tensor process unit}{tpu}

\myacronymentry{HAL}{HAL}{hardware abstraction layer}{hal}

\myacronymentry{SRM}{SRM}{spike response model}{srm}

\myacronymentry{STBP}{STBP}{spatio-temporal back propagation}{stbp}

\myacronymentry{STDP}{SDBP}{spike time dependent plasticity}{stdp}

\myacronymentry{API}{API}{application programming interface}{api}

\myacronymentry{SLAM}{SLAM}{simultaneous localization and mapping}{slam}

\myacronymentry{FISTA}{FISTA}{fast iterative shrinkage-thresholding algorithm}{fista}

\myacronymentry{LASSO}{LASSO}{least absolute shrinkage and selection operator}{lasso}

\myacronymentry{AI}{AI}{artificial intelligence}{ai}

\myacronymentry{ISA}{ISA}{instruction set architecture}{isa}
\newcommand{\ISA}{\gls{ISA}\xspace}

\myacronymentry{IMU}{IMU}{inertial measurement unit}{imu}

\myacronymentry{DVFS}{DVFS}{dynamic voltage and frequency scaling}{dvfs}

\myacronymentry{ABB}{ABB}{adaptive body-biasing}{abb}

\myacronymentry{ULP}{ULP}{ultra-low power}{ulp}

\myacronymentry{ML}{ML}{machine learning}{ml}
\newcommand{\ML}{\gls{ML}\xspace}

\myacronymentry{EHWPE}{EHWPE}{external hardware processing engine}{ehwpe}
\newcommand{\EHWPE}{\gls{EHWPE}\xspace}

\myacronymentry{FC}{FC}{fabric controller}{fc}
\newcommand{\FC}{\gls{FC}\xspace}

\myacronymentry{TEI}{TEI}{temperature effect inversion}{tei}

\myacronymentry{TCDM}{TCDM}{tightly coupled data memory}{tcdm}
\newcommand{\TCDM}{\gls{TCDM}\xspace}

\myacronymentry{UART}{UART}{universal asynchronous receiver transmitter}{uart}

\myacronymentry{CPI}{CPI}{camera parallel interface}{cpi}

\myacronymentry{GPIO}{GPIO}{general purpose input output}{gpio}

\myacronymentry{FLL}{FLL}{frequency locked loop}{fll}
\newcommand{\FLL}{\gls{FLL}\xspace}

\myacronymentry{PWM}{PWM}{pulse-width modulation}{pwm}

\myacronymentry{APB}{APB}{advanced peripheral bus}{apb}
\newcommand{\APB}{\gls{APB}\xspace}

\myacronymentry{JTAG}{JTAG}{joint test action group}{jtag}

\myacronymentry{FPU}{FPU}{floating point unit}{fpu}
\newcommand{\FPU}{\gls{FPU}\xspace}

\myacronymentry{PMU}{PMU}{power management unit}{pmu}

\myacronymentry{PDK}{PDK}{process development kit}{pdk}

\myacronymentry{ALU}{ALU}{arithmetic logic unit}{alu}

\myacronymentry{MAC}{MAC}{multiply-accumulate}{mac}

\myacronymentry{NSAA}{NSAA}{near sensor analytic application}{nsaa}

\myacronymentry{UAV}{UAV}{unmanned aerial vehicle}{uav}

\myacronymentry{kws}{KWS}{keyword spotting}{kws}
\myacronymentry{cnn}{CNN}{convolutional neural network}{cnn}
\myacronymentry{mlp}{MLP}{multi-layer perceptron}{mlp}
